\documentclass[twocolumn,prb]{revtex4}
\usepackage{amsfonts}
\usepackage[T1]{fontenc}
\usepackage{amsmath,amsbsy,amssymb,graphicx}
\usepackage{times}

\let\mathbf=\boldsymbol
\def\AppNum#1{\def\theequation{#1.\arabic{equation}}
\setcounter{equation}{0}}

\begin{document}

\title{Dirac formulation for universal quantum gates and Shor's integer
factorization \\
in high-frequency electric circuits }
\author{Motohiko Ezawa}
\affiliation{Department of Applied Physics, University of Tokyo, Hongo 7-3-1, 113-8656,
Japan}

\begin{abstract}
Quantum computation may well be performed with the use of electric circuits.
Especially, the Schr\"{o}dinger equation can be simulated by the
lumped-element model of transmission lines, which is applicable to
low-frequency electric circuits. In this paper, we show that the Dirac
equation is simulated by the distributed-element model, which is applicable
to high-frequency electric circuits. Then, a set of universal quantum gates
(the Hadamard, phase-shift and CNOT gates) are constructed by networks made
of transmission lines. We demonstrate Shor's prime factorization based on
electric circuits. It will be possible to simulate any quantum algorithms
simply by designing networks of metallic wires.
\end{abstract}

\maketitle

\section{Introduction}

Quantum computation is one of the hottest topic in physics\cite{Feynman,DiVi}. 
Various proposals have been made based on superconducting qubits\cite%
{Nakamura}, ion trap\cite{Cirac}, photonic system\cite{Knill}, quantum dots\cite{Loss} 
and nuclear magnetic resonance\cite{Vander,Kane}. For universal
computations, it is enough to construct only three unitary gates, the
Hadamard, phase-shift and CNOT gates, where all of the unitary gates are
constructed by their combination\cite{Deutsch,Dawson,Universal}. For
instance, a set of universal quantum gates has been constructed based on
quantum walk\cite%
{Child,Varba,Blumer,Hines,Lovett,Webb,MichaelA,MichaelB,Dmitry,Lahini}.
Shor's prime factorization\cite{Shor,Beck,Eckert} has been demonstrated by
using nuclear magnetic resonance\cite{Vander,VanderL}, photonic systems\cite%
{LuShor,Lanyon,Politi,Lopez} and a Josephson junction\cite{Lucero}.

Recently, it was shown that the Schr\"{o}dinger equation is simulated by the
lumped-element model of transmission lines\cite{EzawaSch}. Especially, a set
of universal quantum gate has been constructed solely with the use of $LC$
circuits\cite{EzawaUniv}. This lumped-element model is only valid for
low-frequency electric circuits. It corresponds to the tight-binding model
in the context of condensed matter physics. On the other hand, the
distributed-element model is appropriate for high-frequency electric
circuits. It corresponds to the continuum theory.

In this paper, first we show that the transmission line is described by the
one-dimensional Dirac equation, where the voltage and the current form a
two-component wave function. Next, we construct a set of universal quantum
gates consisting of the Hadamard, phase-shift and CNOT gates. Based on these
gates, we make a demonstration of Shor's prime factorization. Our results
will open a way to simulate quantum algorithms based on distributed-element
electric circuits.

\begin{figure*}[t]
\centerline{\includegraphics[width=0.80\textwidth]{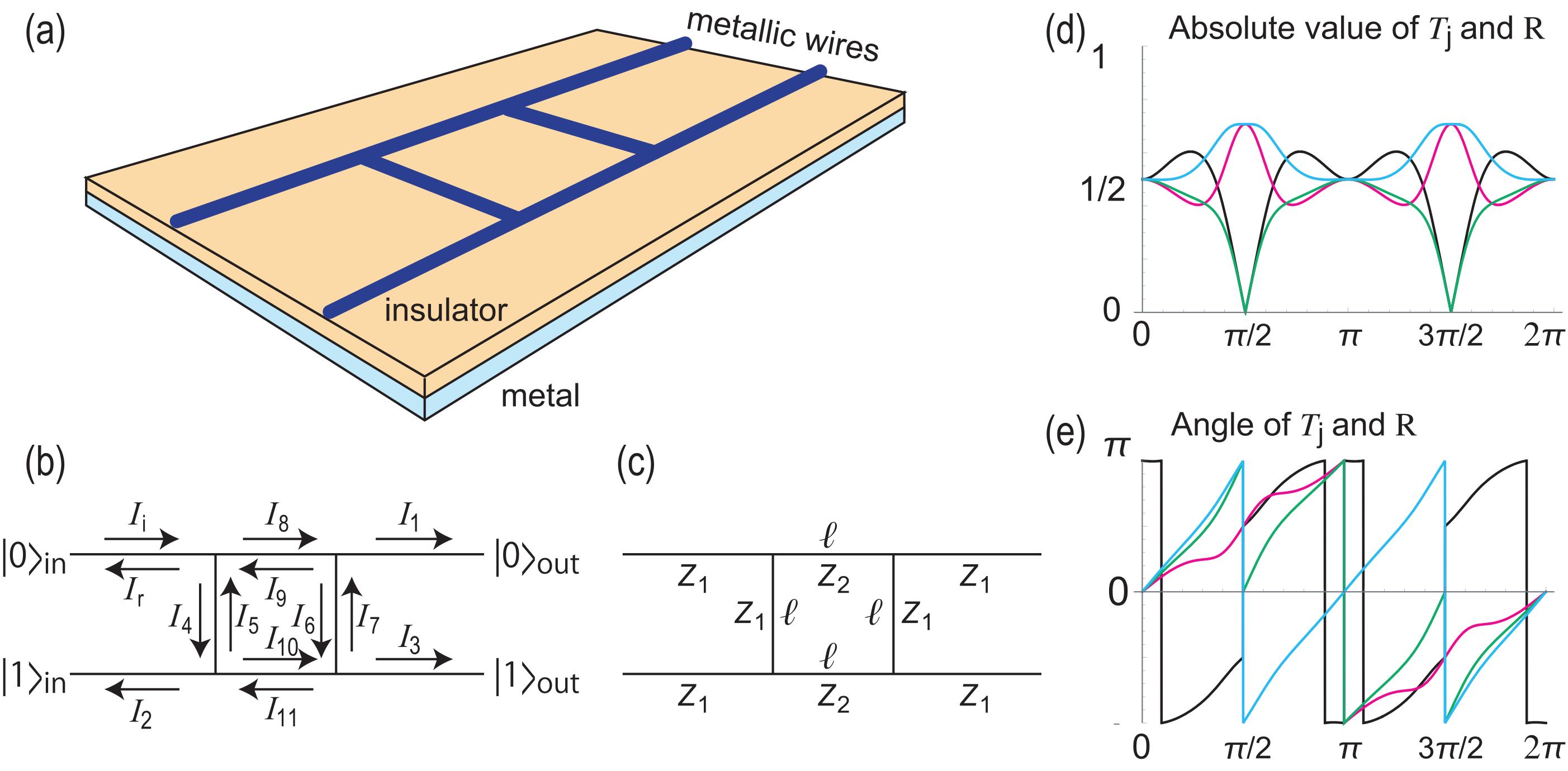}}
\caption{(a) Illustration of a bilayer system made of a metallic plane and
an insulator plane, upon which metallic wires are placed. This figure is for
a mixing gate. (b) and (c) Current $I_{i}$ and impedance $Z_{i}$ with the
index $i$ are used in Appendix B. (d) and (e) The $k$ dependence of the
absolute value and the phase of the transmission coefficients $T_{1}(k)$, 
$T_{2}(k)$, $T_{3}(k)$ and the reflection coefficient $R(k)$. $T_{1}$ is
colored in red, $T_{2}$ is colored in green, $T_{3}$ is colored in cyan and 
$R$ is colored in black. The horizontal axis is the momentum $0\leq k\leq 2%
\protect\pi $. Is it observed that $T_{2}(k)=R(k)=0$ at $k\ell =\protect\pi %
/2$ and $3\protect\pi /2$. We have set $Z_{2}/Z_{1}=1/\protect\sqrt{2}$.}
\label{FigDiracModel}
\end{figure*}

\section{Transmission line and the Dirac equation}

The electrodynamics along a transmission line is governed by the telegrapher
equation made of%
\begin{align}
L\frac{d}{dt}I\left( x,t\right) =& -\frac{\partial }{\partial x}V\left(
x,t\right) ,  \label{EqB} \\
C\frac{d}{dt}V\left( x,t\right) =& -\frac{\partial }{\partial x}I\left(
x,t\right) .  \label{EqA}
\end{align}%
The first equation is the Kirchhoff voltage law, describing the voltage drop
by the self-inductive electromotive force. The second equation is the
Kirchhoff current law.

We consider a bilayer system made of an insulator placed upon a metal: See
Fig.\ref{FigDiracModel}. Metallic wires deposited on this insulator plane
are described by the telegrapher equations (\ref{EqB}) and (\ref{EqA}) with%
\begin{eqnarray}
C &=&\frac{2\pi \varepsilon }{\log \left[ 2h/r\right] }\left[ \text{F/m}\right] , \\
L &=&\frac{\mu }{2\pi }\log \frac{2h}{r}\left[ \text{H/m}\right] ,
\end{eqnarray}%
where $r$ is the radius of the metallic wire, and $h$ is the distance
between the wire and the metallic plane. The capacitive effect between the
wire and the metallic plane leads to the capacitance $C$, while the origin
of the inductance $L$ is the self-inductive electromotive force.

The set of equations (\ref{EqB}) and (\ref{EqA}) are reformulated in the
form of the one-dimensional Dirac equation, 
\begin{equation}
i\partial _{t}\psi \left( x,t\right) =\mathcal{H}\psi \left( x,t\right) ,
\label{DiracEq}
\end{equation}%
with the wave function 
\begin{equation}
\psi \left( x,t\right) =\left( 
\begin{array}{c}
ZI(x,t) \\ 
V\left( x,t\right)%
\end{array}%
\right) ,
\end{equation}%
where $Z=\sqrt{L/C}$ is the characteristic impedance of the wire. The
Hamiltonian is given by 
\begin{equation}
\mathcal{H}=-\left( 
\begin{array}{cc}
0 & \frac{i}{\sqrt{LC}}\frac{\partial }{\partial x} \\ 
\frac{i}{\sqrt{LC}}\frac{\partial }{\partial x} & 0%
\end{array}%
\right) =-\frac{i}{\sqrt{LC}}\sigma _{x}\frac{\partial }{\partial x}.
\end{equation}%
In the momentum space, it is reduced to%
\begin{equation}
\mathcal{H}=\frac{1}{\sqrt{LC}}\sigma _{x}k,
\end{equation}%
whose eigenvalue is given by%
\begin{equation}
E=\pm \frac{1}{\sqrt{LC}}\left\vert k\right\vert ,
\end{equation}%
where $k$ is the momentum. Its solution is a plane wave%
\begin{equation}
\psi \left( x,t\right) =c_{\text{i}}e^{i\left( \omega t-kx\right) }+c_{\text{r}}e^{i\left( \omega t+kx\right) },  \label{PlaneWave}
\end{equation}%
where coefficients $c_{\text{i}}$ and $c_{\text{r}}$ are to be determined by
the boundary conditions. Here the indices "i" and "r" stand for "injected"
and "reflected", respectively.

The total energy $U_{\text{T}}=U_{\text{E}}+U_{\text{M}}$ is conserved along
the transmission line, where 
\begin{equation}
U_{\text{E}}=\frac{C}{2}\sum V^{2},\qquad U_{\text{M}}=\frac{L}{2}\sum I^{2}
\end{equation}%
are the electrostatic energy and the magnetic energy, respectively. On the
other hand, the probability of the wave function is rewritten in the form 
\begin{equation}
\sum \left\vert \psi \right\vert ^{2}=\sum \mathcal{I}^{2}+\mathcal{V}^{2}
=\sum \frac{L}{C}I^{2}+V^{2}=\frac{2}{C}U_{\text{T}}.
\end{equation}%
Hence, the conservation of the probability of the wave function is assured
by the conservation of the total energy. This holds for a generic network
made of several transmission lines.

\section{Quantum Gates}

\textbf{One-qubit gates. }A one-qubit gate $U$ from the input $(\left\vert
0\right\rangle _{\text{in}},\left\vert 1\right\rangle _{\text{in}})$ to the
output $(\left\vert 0\right\rangle _{\text{out}},\left\vert 1\right\rangle _{\text{out}})$ is defined by%
\begin{equation}
\left( 
\begin{array}{c}
\left\vert 0\right\rangle _{\text{out}} \\ 
\left\vert 1\right\rangle _{\text{out}}%
\end{array}%
\right) =U\left( 
\begin{array}{c}
\left\vert 0\right\rangle _{\text{in}} \\ 
\left\vert 1\right\rangle _{\text{in}}%
\end{array}%
\right) .  \label{Qubit}
\end{equation}%
In order to realize them, we use a two-port network of electric circuits
with two inputs and two output as in Fig.\ref{FigDiracModel}.

Linear electric circuits satisfy the superposition principle. We calculate
the transmission and reflection coefficients when we input a plane wave only
to the wire $\left\vert 0\right\rangle _{\text{in}}$. There are three other
lines, where two of them are the outputs and the rest is the other input.

\begin{figure*}[t]
\centerline{\includegraphics[width=0.88\textwidth]{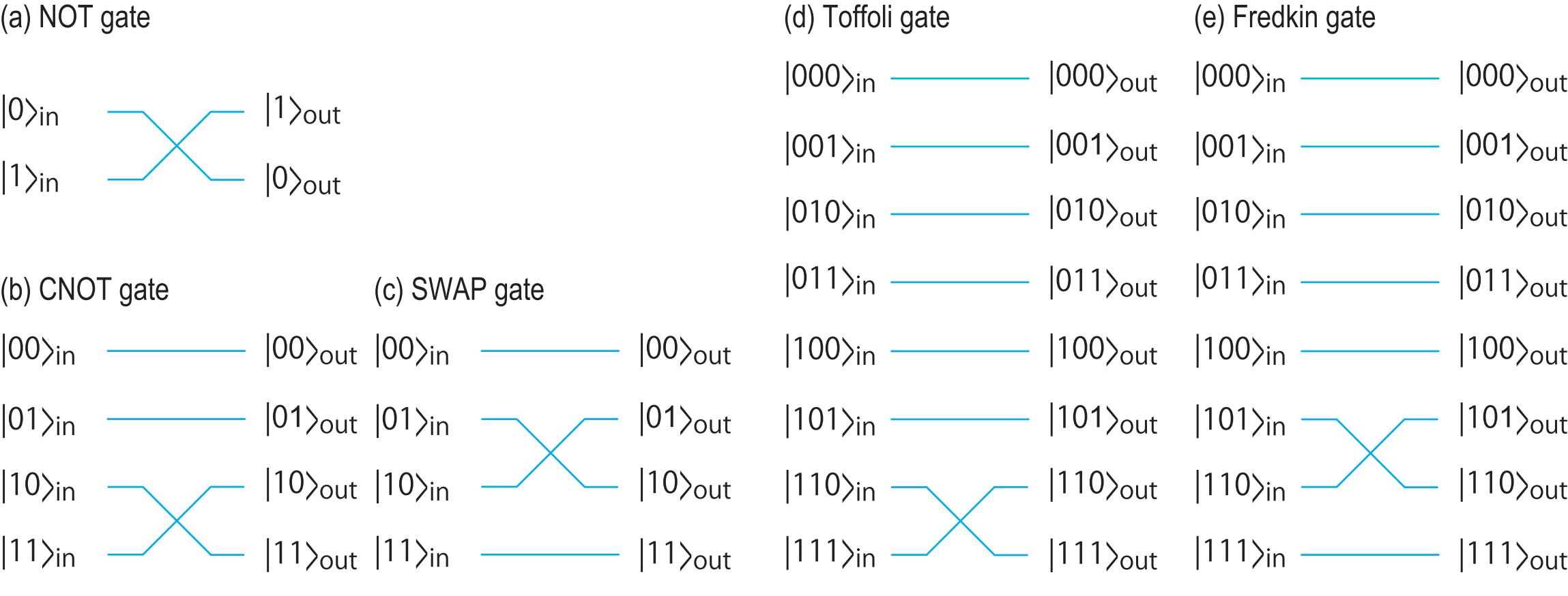}}
\caption{Illustration of classical gates, where some wires are interchanged.
(a) X (NOT) gate, (b) CNOT gate, (c) SWAP gate, (d) Toffoli gate and (e)
Fredkin gate.}
\label{FigToffoli}
\end{figure*}

This is a scattering problem, and the wave functions are written in the form of%
\begin{align}
\left\langle x,0|\psi \right\rangle & =e^{-ikx}+R\left( k\right) e^{ikx}, \\
\left\langle x,j|\psi \right\rangle & =T_{j}\left( k\right) e^{ikx},\qquad
j=1,2,3.
\end{align}%
In general, there is a reflection to the input. It is necessary to tune the
parameters so as to cancel the reflection exactly, which we refer to as the
no-reflection condition.

\textbf{Mixing gate.} As the first example of one-qubit gate, we study a
double bridge structure shown in Fig.\ref{FigDiracModel}, where two inputs
and two output wires are attached to a square with length $\ell $. The
current-voltage relation is $V_{i}=Z_{i}I_{i}$, with $Z_{i}$ being the
characteristic impedance of the wire. The no-reflection condition is given
by 
\begin{equation}
I_{\text{r}}=I_{2}=0.
\end{equation}%
This condition leads to the impedance-matching relation 
\begin{equation}
Z_{2}/Z_{1}=1/\sqrt{2},  \label{ImpMat}
\end{equation}%
and the condition on the length $\ell $%
\begin{equation}
k\ell =\pi /2\quad \text{or}\quad 3\pi /2,  \label{Condi-k-ell}
\end{equation}%
as we derive in Appendix B: See (\ref{step3}) and (\ref{step2}). We show the
transmission and reflection coefficients as a function of $k$ and in Fig.\ref{FigDiracModel}(d) and (e). 
It is observed that $T_{2}(k)=R\left( k\right)=0 $ at $k\ell =\pi /2$ and $k\ell =3\pi /2$

Let us choose $k\ell =\pi /2$. Then, the transmission currents are obtained
as%
\begin{equation}
I_{1}/I_{\text{i}}=i/\sqrt{2},\quad I_{3}/I_{\text{i}}=-1/\sqrt{2}.
\end{equation}%
We compare these with the definition of one-qubit gate (\ref{Qubit}) to find
that%
\begin{equation}
U_{\text{mix}}=\frac{1}{\sqrt{2}}\left( 
\begin{array}{cc}
i & -1 \\ 
-1 & i%
\end{array}%
\right) ,  \label{MixGate}
\end{equation}%
which is the mixing gate.

\textbf{Phase-shift gate. }As the second example of one-qubit gate, we study
a phase-shift gate defined by 
\begin{equation}
U_{\phi }=\left( 
\begin{array}{cc}
1 & 0 \\ 
0 & e^{i\phi }%
\end{array}%
\right) .  \label{PhaseShift}
\end{equation}%
It follows from (\ref{PlaneWave}) that the phase shift is a function of the
length of a wire. Indeed, when the length of upper (lower) wire is $\ell
_{1} $ ($\ell _{2}$) in a two-port network without any interaction between
two wires, the phase shift is given by $e^{ik\ell _{1}}$ ($e^{ik\ell _{2}}$). 
It acts as a quantum gate,%
\begin{equation}
\left( 
\begin{array}{cc}
e^{ik\ell _{1}} & 0 \\ 
0 & e^{ik\ell _{2}}%
\end{array}%
\right) =e^{ik\ell _{1}}\left( 
\begin{array}{cc}
1 & 0 \\ 
0 & e^{ik\left( \ell _{2}-\ell _{1}\right) }%
\end{array}%
\right) .
\end{equation}%
A phase delay is found to occur for an elongated wire. Since the overall
phase is meaningless, the phase shift is given by $\phi =k\left( \ell
_{2}-\ell _{1}\right) $ in (\ref{PhaseShift}). We can construct a
phase-shift gate (\ref{PhaseShift}) with an arbitrary phase by tuning the
length of the elongated wire continuously. This is a merit comparing with
the previous result in the lumped-electric circuit\cite{EzawaUniv}. By tuning%
\emph{\ }$k\left( \ell _{2}-\ell _{1}\right) =\pi $, we can construct a
Pauli Z gate $e^{ik\ell _{1}}\sigma _{Z}$.

\textbf{Hadamard gate. }The Hadamard gate is defined by%
\begin{equation}
U_{\text{H}}=\frac{1}{\sqrt{2}}\left( 
\begin{array}{cc}
1 & 1 \\ 
1 & -1%
\end{array}%
\right) .
\end{equation}%
It is constructed by the combination of the mixing gate and the $3\pi /2$
phase-shift gate as $U_{\text{H}}=-iU_{3\pi /2}U_{\text{mix}}U_{3\pi /2}$.

\textbf{NOT gate. }The NOT gate $U_{X}$ is given by the Pauli $\sigma _{x}$
matrix, whose network is illustrated in Fig.\ref{FigToffoli}(a). It is
constructed by interchanging the labels of the $|0\rangle $ and $|1\rangle $, 
as shown in Fig.\ref{FigToffoli}(b).

\textbf{Two-qubit gates. }We proceed to consider the four-port network,%
\begin{equation}
\left( 
\begin{array}{c}
\left\vert 00\right\rangle _{\text{out}} \\ 
\left\vert 01\right\rangle _{\text{out}} \\ 
\left\vert 10\right\rangle _{\text{out}} \\ 
\left\vert 11\right\rangle _{\text{out}}%
\end{array}%
\right) =U\left( 
\begin{array}{c}
\left\vert 00\right\rangle _{\text{in}} \\ 
\left\vert 01\right\rangle _{\text{in}} \\ 
\left\vert 10\right\rangle _{\text{in}} \\ 
\left\vert 11\right\rangle _{\text{in}}%
\end{array}%
\right) .
\end{equation}%
The most well-known one is the CNOT gate defined by%
\begin{equation}
U_{\text{CNOT}}=\left( 
\begin{array}{cc}
I_{2} & O_{2} \\ 
O_{2} & U_{X}%
\end{array}%
\right) ,
\end{equation}%
where $I_{2}$ is the two-dimensional identity matrix, $O_{2}$ is the
two-dimensional null matrix, and $U_{X}=\sigma _{x}$ is the NOT gate. We
interchange the wires\cite{Child,EzawaUniv} for the states $\left\vert
10\right\rangle $ and $\left\vert 11\right\rangle $, while we keep the
states $\left\vert 00\right\rangle $ and $\left\vert 01\right\rangle $ as
shown in Fig.\ref{FigToffoli}(b). Let the length of the wires for 
$\left\vert 00\right\rangle $ and $\left\vert 01\right\rangle $ to be 
$\ell_{1}$, and that of the wires for $\left\vert 10\right\rangle $ and 
$\left\vert 11\right\rangle $ to be $\ell _{2}$. Although $\ell _{2}\neq \ell_{1}$, 
it is possible to suppress a phase shift between these two types of
wires by setting $k\left( \ell _{2}-\ell _{1}\right) =2\pi $.

We similarly construct the SWAP gate by exchanging the wires $\left\vert
10\right\rangle $\ and $\left\vert 01\right\rangle $ as in Fig.\ref%
{FigToffoli}(c).

\textbf{Three-qubit gates. }It is straightforward to construct three-qubits
gates including the Toffoli and Fredkin gates. We exchange wires between 
$\left\vert 110\right\rangle $ and $\left\vert 111\right\rangle $ in the
Toffoli gate as in Fig.\ref{FigToffoli}(d), while we exchange wires between 
$\left\vert 101\right\rangle $ and $\left\vert 110\right\rangle $ in the
Fredkin gate as in Fig.\ref{FigToffoli}(e).

\begin{figure}[t]
\centerline{\includegraphics[width=0.48\textwidth]{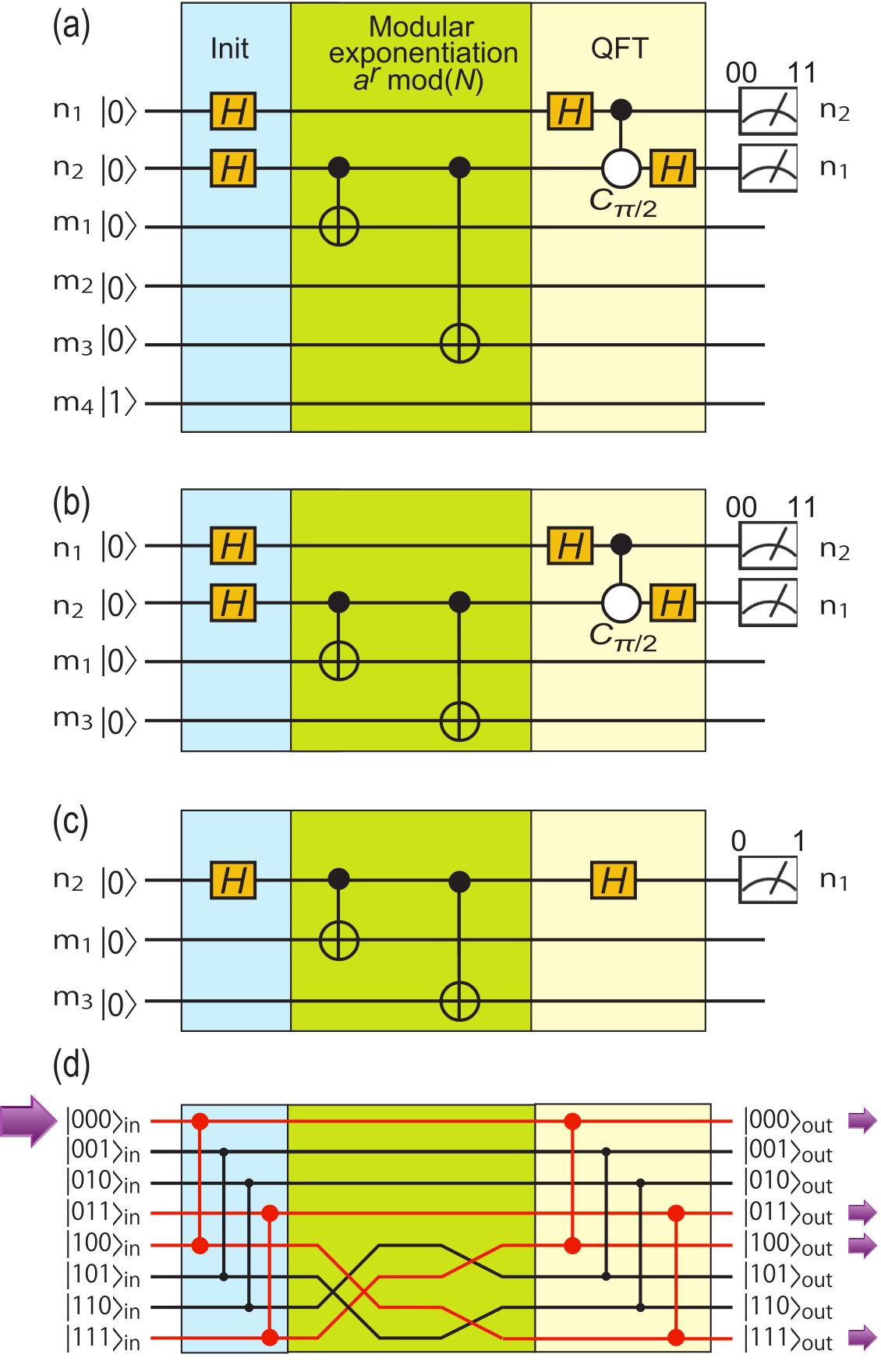}}
\caption{(a) - (c) Quantum circuit\protect\cite{LuShor,Lucero,Bus} for the
period-finding routine of Shor's algorithm with $N=15$ and $a=11$. We note
that $n_{1}$\ and $n_{2}$\ have been swapped in the output. Qubits $m_{2}$
and $m_{4}$ can be deleted since there is no action on these qubits, and (a)
is simplified into (b). Since $U_{\text{H}}^{2}=1$, the qubit $m_{1}$ can be
deleted, and (b) is simplified into (c). (d) Illustration of the
corresponding electric circuit. The vertical line indicates the Hadamard
gate. The current flows only in the red lines.}
\label{FigShor}
\end{figure}

\section{Shor's integer factorization}

As a demonstration we show how to perform Shor's integer factorization in
electric circuits. Shor's algorithm is composed of quantum and classical
parts. The quantum part is a period-finding algorithm, which consists of the
Hadamard gate, the modular exponentiation and the inverse quantum Fourier
transformation (QFT) as shown in Fig.\ref{FigShor}. We simulate the quantum
part in an electric circuit.

We study a typical example of the factorization of $15$, whose quantum
circuit\cite{LuShor,Lopez} is given in Fig.\ref{FigShor}(a). It consists of
six qubits starting with the $|000001\rangle $,\emph{\ }where the first two
qubits $(n_{1},n_{2})$ are called the register qubits while the last four
qubits $(m_{1},m_{2},m_{3},m_{4})$ are called the ancilla qubits. To
simplify the calculation, we make a compilation of this quantum circuit\cite{LuShor,Lopez}. 
First, since there are no actions for the fourth and six
qubits, they can be removed, and we obtain Fig.\ref{FigShor}(b). Next, the
first qubit can be removed since we have $U_{\text{H}}^{2}=1$ and there is
no action for the controlled phase-shift gate since the input is zero. By
removing the first qubit, we have a compiled quantum circuit for three
qubits shown in Fig.\ref{FigShor}(c). We implement it in the electric
circuit as shown in Fig.\ref{FigShor}(d).

We start with a top most wire corresponding to $|000\rangle $. The output
can be read out by measuring the magnitude and the phase of the current for
each wire. We would observe that the magnitudes of currents are identical
for four wires $|000\rangle $, $|011\rangle $, $|100\rangle $ and 
$|111\rangle $ but the phase is different by 180 degree only for $|111\rangle $. 
Then the output is given by%
\begin{equation}
\frac{1}{2}\left( |000\rangle +|011\rangle +|100\rangle -|111\rangle \right)
.  \label{ShorOutput}
\end{equation}%
It is identical\cite{LuShor,Lopez} to the output for the quantum circuit for
the period-finding routine of Shor's algorithm.

The result (\ref{ShorOutput}) is the one in the compiled circuit. By
recovering the removed qubits, the output reads in the full quantum circuit
as 
\begin{equation}
\frac{1}{2}\left( |000001\rangle +|100001\rangle +|001011\rangle
-|101011\rangle \right) .  \label{FullOutput}
\end{equation}%
The compilation of (\ref{FullOutput}) to (\ref{ShorOutput}) is understood by
noting that the second qubit (0), the fourth qubit (0) and the sixth qubit
(1) are common for all four terms in (\ref{FullOutput}). Namely, since there
is no action for the second, fourth and sixth qubits, there is no need to
apply unitary transformation in the quantum circuit. Here we note that 
$n_{1} $\ and $n_{2}$\ have been swapped after the QFT\cite{Universal}. It
follows from (\ref{FullOutput}) that the register qubits are $|00\rangle $
and $|10\rangle $. As reviewed in Appendix C, we find the period $r=2$ from
this output, and we obtain the prime factorization $15=3\times 5$.

\section{Conclusion}

We have constructed a set of universal quantum gates based on the
distributed-element model applicable to high-frequency electric circuits. We
can construct them only by using metallic wires deposited on an insulating
layer placed on the metallic layer. The size of the system will be greatly
reduced to the order of 10nm. Our results will open a way for integrated
circuits for simulating quantum algorithms.

\section*{Acknowledgement}

The author is very much grateful to A. Kurobe, E. Saito and N. Nagaosa for helpful
discussions on the subject. This work is supported by the Grants-in-Aid for
Scientific Research from MEXT KAKENHI (Grants No. JP17K05490 and No.
JP18H03676). This work is also supported by CREST, JST (JPMJCR16F1).

\begin{figure}[t]
\centerline{\includegraphics[width=0.3\textwidth]{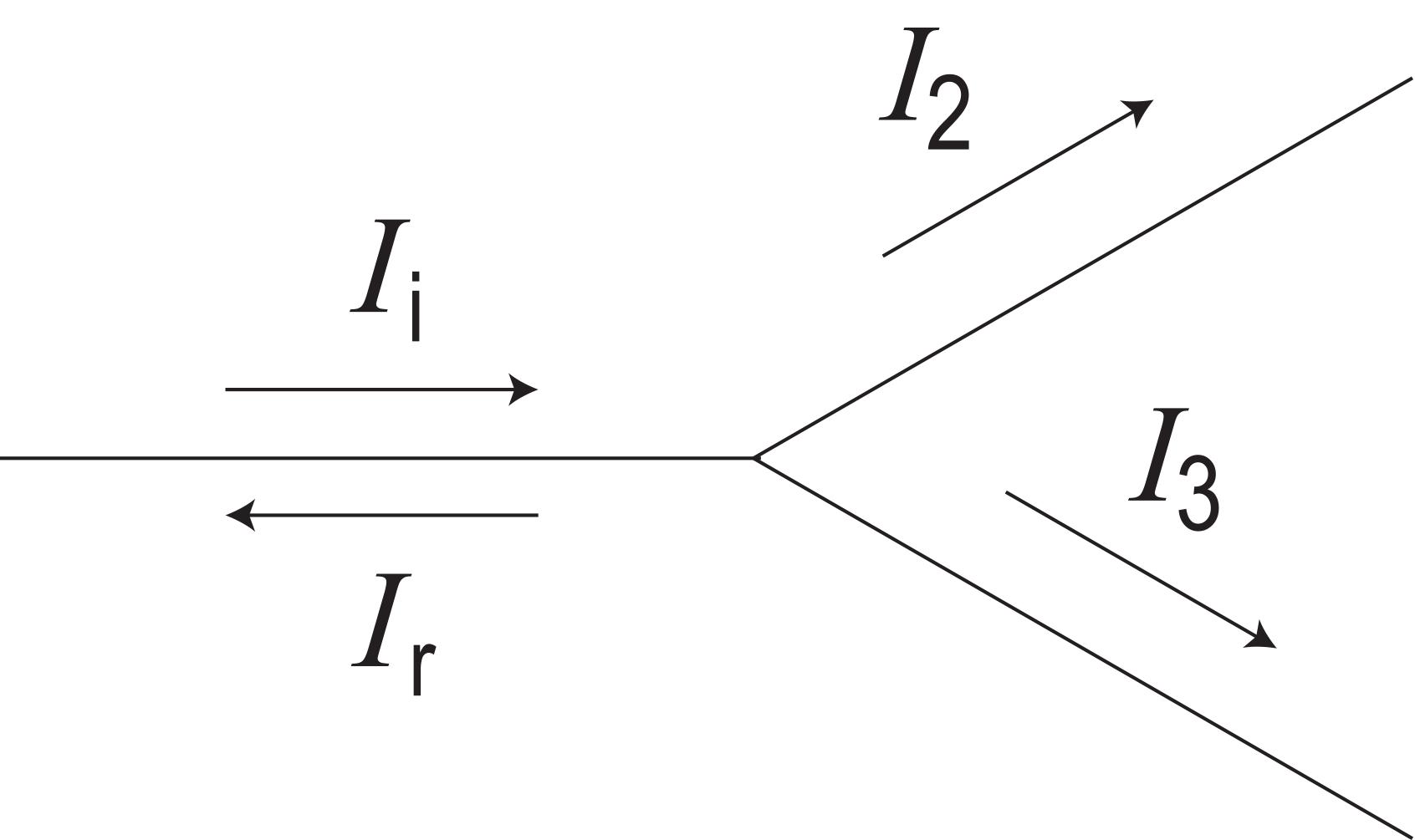}}
\caption{ (a) Illustration of the Y junction. The numbers $i$ show the index
for $I_{i}$.}
\label{FigYJunc}
\end{figure}

\AppNum{A}

\section*{Appendix A: Y-junction\textbf{\ }}

We study a transmission through a Y-junction (Fig.\ref{FigYJunc}). There are
three legs, which we call the leg-$i$. We inject a current $I_{\text{i}}$\
to the leg-$1$. There is a reflected current $I_{\text{r}}$\ by the junction
in general. Thus, two currents flow on the leg-$1$, with the total current
being $I_{\text{i}}-I_{\text{r}}$. Let $Z_{i}$\ be the characteristic
impedance of the leg-$i$. The voltage $V_{i}$ and the current $I_{i}$ are
related by the impedance $Z_{i}$ as%
\begin{equation}
V_{\text{i}}=Z_{1}I_{\text{i}},\quad V_{\text{r}}=Z_{1}I_{\text{r}},\quad
V_{2}=Z_{2}I_{2},\quad V_{3}=Z_{3}I_{3}.  \label{Y1}
\end{equation}%
At the junction, the current conservation gives%
\begin{equation}
I_{\text{i}}-I_{\text{r}}=I_{2}+I_{3},  \label{Y3}
\end{equation}%
while the voltages are related as%
\begin{equation}
V_{\text{i}}+V_{\text{r}}=V_{2}=V_{3}.  \label{Y2}
\end{equation}%
It follows from (\ref{Y1}), (\ref{Y2}) and (\ref{Y3}) that 
\begin{align}
V_{\text{r}}& =\frac{Z_{2}Z_{3}-Z_{1}\left( Z_{2}+Z_{3}\right) }{%
Z_{1}Z_{2}+Z_{2}Z_{3}+Z_{3}Z_{1}},  \label{Y4} \\
I_{\text{r}}& =-\frac{Z_{2}Z_{3}-Z_{1}\left( Z_{2}+Z_{3}\right) }{%
Z_{1}\left( Z_{1}Z_{2}+Z_{2}Z_{3}+Z_{3}Z_{1}\right) }.  \label{Y5}
\end{align}%
We require no reflection at the junction, which implies%
\begin{equation}
V_{\text{r}}=I_{\text{r}}=0.  \label{Y6}
\end{equation}%
Solving (\ref{Y4}) and (\ref{Y5}) with (\ref{Y6}), we obtain%
\begin{equation}
Z_{1}=\frac{Z_{2}Z_{3}}{Z_{2}+Z_{3}},  \label{Z1}
\end{equation}%
which is the impedance matching condition. Consequently, we obtain 
\begin{align}
V_{2}& =Z_{1}I_{\text{i}},\qquad V_{3}=Z_{1}I_{\text{i}}, \\
I_{2}& =\frac{Z_{3}}{Z_{2}+Z_{3}}I_{\text{i}},\qquad I_{3}=\frac{Z_{2}}{%
Z_{2}+Z_{3}}I_{\text{i}}
\end{align}%
for the transmissions along the leg-$2$ and the leg-$3$.

\AppNum{B}

\section*{Appendix B: Mixing gate\textbf{\ }}

We study a transmission through the mixing gate (Fig.\ref{FigDiracModel}),
where two inputs and two output wires are attached to a square with length $\ell $. 
By using the notation for the currents in Fig.\ref{FigDiracModel}(b), 
the current conservations give%
\begin{align}
& I_{\text{i}}-I_{\text{r}}-I_{4}+I_{5}-I_{8}+I_{9}=0, \\
& -I_{2}+I_{4}e^{ik\ell }-I_{5}e^{-ik\ell }-I_{10}+I_{11}=0, \\
& -I_{1}-I_{6}+I_{7}+I_{8}e^{ik\ell }-I_{9}e^{-ik\ell }=0, \\
& -I_{3}+I_{6}e^{ik\ell }-I_{7}e^{-ik\ell }+I_{10}e^{ik\ell
}-I_{11}e^{-ik\ell }=0,
\end{align}%
while the voltage relations are%
\begin{align}
& V_{\text{i}}+V_{\text{r}}=V_{4}+V_{5}=V_{8}+V_{9}, \\
& V_{1}=V_{8}e^{ik\ell }+V_{9}e^{-ik\ell }=V_{6}+V_{7}, \\
& V_{2}=V_{4}e^{ik\ell }+V_{5}e^{-ik\ell }=V_{10}+V_{11}, \\
& V_{3}=V_{6}e^{ik\ell }+V_{7}e^{-ik\ell }=V_{10}e^{ik\ell
}+V_{11}e^{-ik\ell }.
\end{align}%
The current-voltage relation reads%
\begin{equation}
V_{i}=Z_{i}I_{i},
\end{equation}%
with $Z_{i}$ being the characteristic impedance of the wire.

When we assume $Z_{i}=Z_{1}$ for $i=$i$,$r$,1,\cdots 7$ and $Z_{i}=Z_{2}$
for $i=8,\cdots 11$ as shown in Fig.\ref{FigDiracModel}(c), we can solve
these equations as 
\begin{align}
I_{\text{r}}/I_{\text{i}}& =Z_{1}^{2}\frac{2Z_{2}^{2}-Z_{1}^{2}}{%
Z_{1}^{4}+4Z_{2}^{4}},\quad I_{2}/I_{\text{i}}=2iZ_{2}^{2}\frac{%
2Z_{2}^{2}-Z_{1}^{2}}{Z_{1}^{4}+4Z_{2}^{4}},  \label{I1} \\
I_{1}/I_{\text{i}}& =\frac{Z_{1}^{3}Z_{2}}{Z_{1}^{4}+4Z_{2}^{4}},\qquad
I_{3}/I_{\text{i}}=\frac{Z_{1}Z_{2}^{3}}{Z_{1}^{4}+4Z_{2}^{4}},  \label{I2}
\end{align}%
for 
\begin{equation}
k\ell =\pm \pi /2,
\end{equation}%
which is (\ref{Condi-k-ell}) in the text. Imposing the no-reflection
condition ($I_{\text{r}}=I_{2}=0$) on (\ref{I1}), we have the impedance
matching condition 
\begin{equation}
Z_{2}/Z_{1}=1/\sqrt{2},  \label{step3}
\end{equation}%
which is (\ref{ImpMat}) in the text. From (\ref{I2}) we obtain 
\begin{equation}
I_{1}/I_{\text{i}}=\pm i/\sqrt{2},\quad I_{3}/I_{\text{i}}=-1/\sqrt{2}.
\label{step2}
\end{equation}%
By comparing (\ref{step2}) with Fig.\ref{FigDiracModel}(b), it is found to
act as%
\begin{equation}
U_{\text{mix}}=\frac{1}{\sqrt{2}}\left( 
\begin{array}{cc}
\pm i & -1 \\ 
-1 & \pm i%
\end{array}%
\right) ,
\end{equation}%
which is the mixing gate (\ref{MixGate}) for $k\ell =\pi /2$ in the text.

\AppNum{C}

\section*{Appendix C: Shor's algorithm\textbf{\ }}

We review Shor's algorithm\cite{Shor,Beck,Eckert} for prime factorization of
an integer $N$ with the aid of the period $r$. It consists of three steps.
The first step is to design a modular exponentiation part of a quantum
circuit, which is done by a classical computer. The second step to find the
period $r$, which is done by a quantum computer. The final step is to obtain
prime factors from $r$, which is done by a classical computer.

We factorize an integer $N=pq$, with both $p$ and $q$ being odd primes. We
pick a random number $a$ satisfying $0<a<N$, which has no common factor with 
$N$. We define the modular exponential function by%
\begin{equation}
f\left( x\right) =a^{x}\text{ (mod}N\text{)}.
\end{equation}%
The Euler theorem dictates that there is a positive integer $r$ satisfying%
\begin{equation}
f\left( r\right) =a^{r}\text{ (mod}N\text{)}=1.  \label{StepB}
\end{equation}%
There is a periodicity,%
\begin{equation}
f\left( x+r\right) =f\left( x\right) ,  \label{StepC}
\end{equation}%
since%
\begin{equation}
a^{x+r}\text{ (mod}N\text{)}=a^{r}\text{ (mod}N\text{)}.
\end{equation}%
Then, we find%
\begin{equation}
a^{r}-1=cN
\end{equation}%
with an integer $c$. It is rewritten as%
\begin{equation}
\left( a^{r/2}+1\right) \left( a^{r/2}-1\right) =cN.  \label{StepD}
\end{equation}%
If $r$ is an even number, at least one nontrivial factor of $N$ is given by
the greatest common denominator of gcd($a^{r/2}+1,N$) or gcd($a^{r/2}-1,N$).
It is solved by using the Euclidean algorithm, which is efficiently
calculated by a classical computer. If $r$ is an odd number, we rechoose a
different number $a$ and redo the process.

Shor's algorithm provides us with an efficient quantum circuit to find the
period $r$. We initialize the state as%
\begin{equation}
\left( \bigotimes\limits^{n}\left\vert 0\right\rangle \right) \left(
\bigotimes\limits^{m-1}\left\vert 0\right\rangle \right) \otimes \left\vert
1\right\rangle,  \label{StepA}
\end{equation}%
where the first $n$ qubits are the register qubits while the second $m$
qubits are the ancilla qubits. We choose $m$ such that $2^{m-1}<N\leq 2^{m}$%
, and a certain integer $n$ of the order of $m$.

We first apply the Hadamard gates on the register qubits, which transforms
the initialized state as%
\begin{equation}
\bigotimes\limits^{n}\left\vert 0\right\rangle \longmapsto \frac{1}{\sqrt{%
2^{n}}}\left( \left\vert 0\right\rangle +\left\vert 1\right\rangle \right)
^{\otimes n}=\frac{1}{\sqrt{2^{n}}}\sum_{x=0}^{2^{n}-1}\left\vert
x\right\rangle ,
\end{equation}%
where $\left\vert x\right\rangle $ stands for the binary representation of $%
x $. By applying the modular exponentiation to the ancilla qubits, (\ref%
{StepA}) leads to,%
\begin{equation}
\frac{1}{\sqrt{2^{n}}}\sum_{x=0}^{2^{n}-1}\left\vert x\right\rangle
\left\vert a^{x}\text{ (mod}N\text{)}\right\rangle .  \label{ModN}
\end{equation}%
Next, we apply a QFT to the register qubits, obtaining%
\begin{equation}
\frac{1}{2^{n}}\sum_{y=0}^{2^{n}-1}\sum_{x=0}^{2^{n}-1}e^{2\pi
ixy/2^{n}}\left\vert y\right\rangle \left\vert a^{x}\text{ (mod}N\text{)}%
\right\rangle .
\end{equation}%
We reorder this sum as%
\begin{equation}
\frac{1}{2^{n}}\sum_{z=0}^{N-1}\sum_{y=0}^{2^{n}-1}\left[ \sum_{x=\left\{
0,\cdots 2^{n}-1\right\} ;f\left( x\right) =z}^{2^{n}-1}e^{2\pi ixy/2^{n}}%
\right] \left\vert y\right\rangle \left\vert z\right\rangle .
\end{equation}%
Since $x$ is periodic as in (\ref{StepC}), we can write it as 
\begin{equation}
x=x_{0}+rb,
\end{equation}%
with $b$ being an integer. The sum is calculated as%
\begin{align}
\sum_{x=\left\{ 0,\cdots 2^{n}-1\right\} ;f\left( x\right) =z}^{2^{n}-1}&
e^{2\pi ixy/2^{n}}  \notag \\
& =e^{2\pi ix_{0}y/2^{n}}\sum_{b=0}^{m-1}e^{2\pi irby/2^{n}},
\end{align}%
where 
\begin{equation}
m-1=\left\lfloor \frac{2^{n}-x_{0}-1}{r}\right\rfloor
\end{equation}%
with the use of a floor function. The absolute value of the coefficient of
the state $\left\vert y\right\rangle \left\vert z\right\rangle $ is given by%
\begin{equation}
\left\vert \frac{1}{2^{n}}\sum_{b=0}^{m-1}e^{2\pi irby/2^{n}}\right\vert ,
\end{equation}%
where%
\begin{equation}
\frac{1}{2^{n}}\sum_{b=0}^{m-1}e^{2\pi irby/2^{n}}=\left\{ 
\begin{array}{cc}
1 & \text{if }e^{2\pi iry/2^{n}}=1 \\ 
\frac{1}{2^{n}}\frac{e^{2\pi iry}-1}{e^{2\pi iry/2^{n}}-1} & \text{if }%
e^{2\pi iry/2^{n}}\neq 1%
\end{array}%
\right. .
\end{equation}%
Hence the coefficient of the register qubits $\left\vert y\right\rangle $
becomes negligible unless 
\begin{equation}
ry/2^{n}\in \mathbb{Z}.  \label{ry}
\end{equation}
Consequently, $r$ can be determined. The prime factors are given by the
nontrivial greatest common divisor of $a^{r/2}\pm 1$.

We take an example\cite{LuShor,Lopez} of $N=15$. Let us choose $a=11$. We
use $m=4$ for ancilla qubits to satisfy $2^{m-1}<15\leq 2^{m}$. It is enough
to use $n=2$ for the register qubits\cite{LuShor,Lopez}. Then, by
calculating (\ref{ModN}), we find%
\begin{align}
& \frac{1}{2}\sum_{x=0}^{3}\left\vert x\right\rangle \left\vert 11^{x}\text{
(mod}15\text{)}\right\rangle   \notag \\
& =\frac{1}{2}\left( \left\vert 0\right\rangle \left\vert 1\right\rangle
+\left\vert 1\right\rangle \left\vert 11\right\rangle +\left\vert
2\right\rangle \left\vert 1\right\rangle +\left\vert 3\right\rangle
\left\vert 11\right\rangle \right)   \notag \\
& =\frac{1}{2}\left( \left( \left\vert 0\right\rangle +\left\vert
2\right\rangle \right) \left\vert 1\right\rangle +\left( \left\vert
1\right\rangle +\left\vert 3\right\rangle \right) \left\vert 11\right\rangle
\right) .
\end{align}%
In the binary representation, it becomes%
\begin{equation}
\frac{1}{2}\left( \left\vert 00\right\rangle \left\vert 0001\right\rangle
+\left\vert 01\right\rangle \left\vert 1011\right\rangle +\left\vert
10\right\rangle \left\vert 0001\right\rangle +\left\vert 11\right\rangle
\left\vert 1011\right\rangle \right) .  \label{step1}
\end{equation}%
The inverse QFT with respect to the register qubits is explicitly given by%
\begin{equation}
U_{\text{QFT2}}^{-1}=\frac{1}{2}\left( 
\begin{array}{cccc}
1 & 1 & 1 & 1 \\ 
1 & -i & -1 & i \\ 
1 & -1 & 1 & -1 \\ 
1 & i & -1 & -i%
\end{array}%
\right) .
\end{equation}%
After the inverse QFT, (\ref{step1}) becomes 
\begin{align}
& \frac{1}{4}\left( \left\vert 00\right\rangle +\left\vert 01\right\rangle
+\left\vert 10\right\rangle +\left\vert 11\right\rangle \right) \left\vert
0001\right\rangle   \notag \\
& +\frac{1}{4}\left( \left\vert 00\right\rangle -i\left\vert 01\right\rangle
-\left\vert 10\right\rangle +i\left\vert 11\right\rangle \right) \left\vert
1011\right\rangle   \notag \\
& +\frac{1}{4}\left( \left\vert 00\right\rangle -\left\vert 01\right\rangle
+\left\vert 10\right\rangle -\left\vert 11\right\rangle \right) \left\vert
0001\right\rangle   \notag \\
& +\frac{1}{4}\left( \left\vert 00\right\rangle +i\left\vert 01\right\rangle
-\left\vert 10\right\rangle -i\left\vert 11\right\rangle \right) \left\vert
1011\right\rangle   \notag \\
& =\frac{1}{2}\left( \left\vert 00\right\rangle +\left\vert 10\right\rangle
\right) \left\vert 0001\right\rangle +\frac{1}{2}\left( \left\vert
00\right\rangle -\left\vert 10\right\rangle \right) \left\vert
1011\right\rangle   \notag \\
& =\frac{1}{2}\left( |000001\rangle +|100001\rangle +|001011\rangle
-|101011\rangle \right) .  \label{StepE}
\end{align}%
We focus on the register qubits, where there are only two states $\left\vert
00\right\rangle $ and $\left\vert 10\right\rangle $, or $\left\vert
0\right\rangle $ and $\left\vert 2\right\rangle $ in the decimal unit. They
imply $y=0$ and $y=2$. By substituting $y=2$ and $n=2$ in (\ref{ry}), we
find $r=2$.

The prime factorization of $N=15$ is done once we find the period $r=2$ for
the choice of $a=11$. Then,%
\begin{equation}
a^{r/2}+1=12,\qquad a^{r/2}-1=10.
\end{equation}%
The greatest common diviser $c$ is found from (\ref{StepD}) as%
\begin{equation}
c=\left( a^{r/2}+1\right) \left( a^{r/2}-1\right) /N=8.
\end{equation}%
Using the Eucledian algorithm, we have%
\begin{equation}
\text{gcd}(12,15)=3,\qquad \text{gcd}(10,15)=5,
\end{equation}%
and hence that $15=3\times 5$.

\end{document}